\documentclass{cpbtex}

\begin{document}

\title{Long  Distance  Daylight  Drone-based Quantum Key Distribution under Relative Motion}

\author{Chun Zhou$^{1,2}$, Yanyang Zhou$^{1,2,\dagger}$, Ping Wang$^{1,2,\dagger}$,
Tan Li$^{1,2}$, \\ Yu Zhou$^{1,2}$, Hao Wang$^{1,2}$, Yanmei Zhao$^3$, Huan Li$^3$,
\\ Dawei Li$^3$, Fangxiang Wang$^4$, Diyuan Zheng$^{1,2}$, Qifa Zhang$^3$, \\ Hui Sun$^5$,
 Shibiao Tang$^6$, Hongwei Li$^{1,2}$, Zhengfu Han$^{4,*}$, Wansu Bao$^{1,2}$\thanks{Corresponding author. E-mail:zfhan@ustc.edu.cn, bws@qiclab.cn}\\
     $^{1}${Key Laboratory of Quantum Information and Cryptography, Zhengzhou, 450001, China }\\
     $^{2}${Synergetic Innovation Centre of Quantum Information and Quantum Physics, Hefei, 230026, China}\\
     $^{3}${Anhui Asky Quantum Technology Co.Ltd, Wuhu, 241102, China}\\
     $^{4}${University of Science and Technology of China, Hefei, 230026, China}\\
     $^{5}${The 34th Research Institute of CETC, Guiling, 541004, China}\\
     $^{6}${QuantumCTeK Co.Ltd, Hefei, 230088, China}}
\date{}
\maketitle
\begin{abstract}
Low-altitude drones can serve as dynamic nodes apparently mitigating terrain-induced impacts for quantum networks. However, it is extremely hard to establish a sable quantum link in a drone-based dynamic platform, which requires centimeter-level positioning techniques and high-precision time synchronization technologies. In this paper, we develop a single-ended polarization adaptive correction technology at both the transmitting and receiving ends. Based on this, we present the world's first kilometer-scale drone-based QKD network, achieving an 1.2 km free-space QKD link with a secure key rate of 2.76 kbps, suitable for urban quantum network deployment. We validate the feasibility of QKD between dynamic drone and ground unmanned vehicle at a relative speed of 1 m/s over a distance of 100 m, attaining a secure key rate of 70.94 kbps. This work advances drone-based QKD from static demonstrations to practical dynamic network, boasting great development potential for an airborne quantum internet.
\end{abstract}

\section{Introduction}
Quantum communication is advancing from laboratory research toward scaled-up deployment, progressing toward the construction of an integrated space-ground and globally accessible quantum network, as shown in Fig.1a. Achieving long-distance quantum networks with high-speed, daylight operation, and dynamically reconfigurable capabilities stands as the core research objective in the contemporary field of Quantum Key Distribution (QKD). While researchers have successfully demonstrated fiber-based QKD transmission over a distance of 1002km\cite{liu2023experimental}, the large-scale deployment of fiber quantum networks is still plagued by notable infrastructure-related constraints\cite{2025A}. 
As an alternative approach, free-space quantum communication not only overcomes these limitations but also combines the advantages of flexible deployment and wide geographical coverage\cite{2023Unmanned}. For instance, in satellite-based QKD applications, research teams have verified intercontinental key distribution across 12900 km\cite{li2025microsatellite}. Meanwhile, drones, as a supplementary solution capable of effectively compensating for the limitations of satellite-based QKD, such as its restricted satellite coverage footprint and the inherently short duration of orbital passes, have attracted considerable attention from researchers\cite{2017Long,scarinzi2023optimization}. Current technical schemes have successfully implemented quantum-secure key distribution verification under relatively static conditions, such as those encountered by drones\cite{2024Experimental,2025Drone}. It is worth noting that drone and fixed-wing aircraft will serve as cross-domain communication links in the future, enabling full-domain situational awareness and efficient interconnection among land-based command nodes, offshore clean energy facilities, and deep-sea operation platforms, thereby providing support for integrated emergency support and energy management(Fig.\ref{fig1}b) . To realize the application goal, the core operational requirement is establishing  kilometer-scale daytime quantum communication links under relative motion.

The pivotal challenge in constructing kilometer-scale daytime QKD links centers on three primary technical hurdles: background noise due to sunlight suppression,  pointing error calibration, and Doppler shift compensation\cite{2018Free,2021UAV,2010Effect,2021Feasibility}. First, high-intensity background noise severely degrades the signal-to-noise ratio (SNR) of QKD systems and elevates the quantum bit error rate\cite{2023Daylight}. Studies have confirmed that integrating multi-dimensional anti-jamming technologies, including spatial, spectral, and temporal filtering, can achieve efficient suppression of solar background noise\cite{scriminich2022optimal,cai2024free}. Recently, signal sources operating in the 1550 nm band possess inherent daytime anti-jamming capabilities, offering a highly promising technical route for enabling continuous daytime QKD operations\cite{2017Long,avesani2021full,2025Free}. Second, pointing errors directly impair photon detection efficiency at the receiver and the secure key generation rate, hindering the practical implementation of drone-based QKD technology\cite{bhatnagar2014performance,trinh2019effects,alshaer2021reliability,ibrahim2022performance,0Pointing}. In the pointing, acquisition, and tracking (PAT) system, a sub-milliradian ultra-narrow field-of-view (FOV) design or increasing the aperture size of the receiver can effectively reduce link alignment losses\cite{dabiri2025unified,0Pointing}. Third, in the dynamic link scenarios, Doppler shift results in a decreased key generation rate, a reduction in the effective link communication time, and even distribution failure\cite{zhao2010effect}. While conventional approaches predominantly depend on broadband filtering or real-time frequency calibration\cite{wang2021feasibility}, emerging research demonstrates that Doppler-adaptive pulse shaping offers a more efficient suppression of spectral drift and broadening \cite{2025Pulse}. This significantly improves the secure key rate and link robustness in dynamic environments.

In our work, we integrate the continuous-variable QKD (CV-QKD) framework with classical laser communication technology, achieving long-distance daytime quantum communication under relative motion. To address the aforementioned challenges, we adopt three key technical designs:
$\left( i\right)$ We use a 1550 nm laser and continuous polarization encoding. By adopting an on-board local oscillator (LO) scheme with heterodyne detection methods, we simultaneously implement spatial and temporal filtering, effectively suppressing background light noise in daytime environments. $\left( ii\right)$ We reduce inherent pointing errors through a coaxial multi-optical path design. Utilizing high-precision positioning data from the BeiDou Navigation System achieveS rapid acquisition and tracking, together with a dual-beacon structure and a centroid algorithm. Mechanical vibration can cause beam misalignment, thereby leading to signal attenuation and an increase in the bit error rate \cite{salih2024enhancing}. To address this issue, we integrate passive vibration isolation technology and a high-precision inertial navigation unit, further enhancing pointing stability and vibration resistance.
$\left( iii\right)$Our design integrates a quantum signal control chassis with a dedicated synchronization module, achieving system synchronization with an accuracy exceeding 200 ps. We additionally adopt a synchronous frame mechanism to ensure the precise coordinated transmission of quantum signals.

Within the urban atmospheric channel, our system successfully suppressed composite interference including high back noise, airframe vibration, and Doppler shift. We established a hundred-meter-level QKD link between the drone and ground nodes during various dynamic scenarios such as hovering at different altitudes and performing return flights, and realize a drone-to-fixed-point link with a distance of 1.2 km. Specifically, the average secure key rate reached 79.48 kbps between the drone hovering at an altitude of 75 m and the unmanned ground vehicle (UGV) with a horizontal distance of 100 m. At the same horizontal distance, the average secure key rate between the slowly flying drone and the UGV reached 70.94 kbps. Meanwhile, the average secure key rate between the hovering drone and the fixed point with a distance of 1.2 km was 2.76 kbps. This marks a key progress in the construction of all-time, kilometer-scale free-space mobile quantum nodes.

\section{RESULTS}\label{sec2}

\subsection{Description of drone to unmanned ground vehicle system }

To achieve high-speed, daytime-compatible, and long-distance drone-based QKD, we illustrate the overall structural design of our experimental setup (including drone terminals and UGV terminals) and their inter-component connections in Fig.1d.
\begin{center}
   \includegraphics{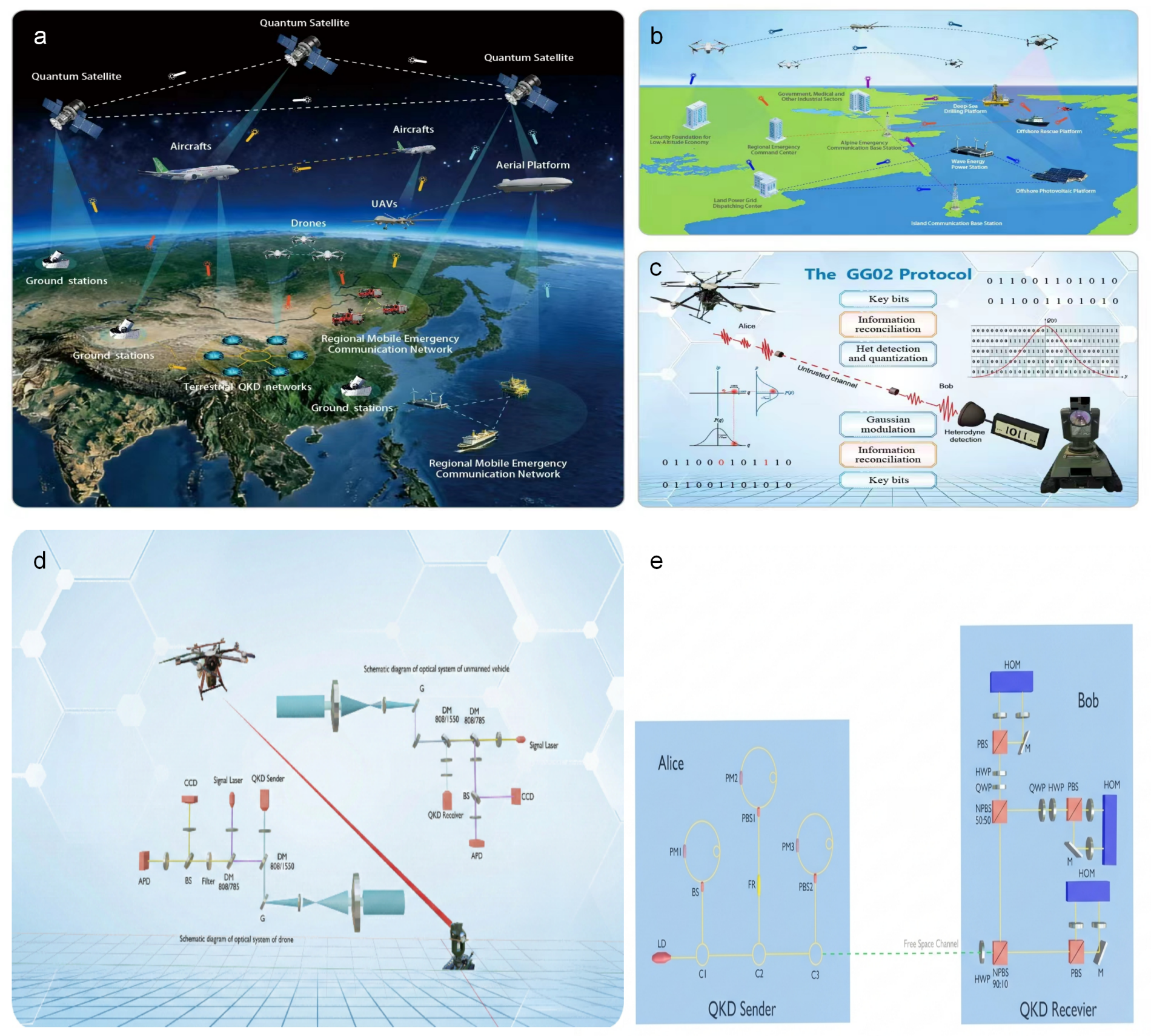}\\[8pt]
	\parbox[c]{15.0cm}{\footnotesize{\bf Fig.~1.}  
		Quantum secured communication network. (a).Space-ground integrated quantum emergency communication network. (b)Drone-based land-sea-air integrated collaborative application scenarios. (c)Scene diagram of free-space continuous-variable quantum key distribution. (d)the schematic diagram of the laser communication system between the drone and the ground unmanned vehicle. (e)Schematic diagram of the CV-QKD optical system.\label{fig1}}
\end{center}
The aerial drone and the UGV on the ground are equipped with a transmitter and a receiver of CV-QKD, respectively, and are connected via a free-space channel. As shown in Fig.1d, the drone's quantum light emission aperture is 50 mm, with an emission optical focal length of approximately 172 mm. It employs a Keplerian telescope optical path featuring a 60 mm receiving aperture. This link comprises a signal light transmission link, a signal light reception link, a precision tracking link, and a quantum light emission link. The UGV has an effective receiving aperture of 246 mm and a system focal length of 260 mm, also adopting a Keplerian telescope optical path with an 250 mm receiving aperture. Its link comprises a signal optical transmission path, a signal optical reception path, a precision tracking path, and a quantum optical reception path. In this setup, a dichroic mirror splits the beam in the main optical path, while optical filters provide wavelength isolation with an extinction ratio of 70 dB.

The tracking system adopted achieves a coarse tracking accuracy of 323 $\mu$rad and a fine tracking accuracy of 38 $\mu$rad. The implementation process of its high-precision tracking function is as follows: Based on Beidou positioning information, the drone emits 785 nm beacon light to scan the area where the UGV is located. After receiving the beacon light, the UGV's coarse acquisition camera performs real-time tracking of it and simultaneously transmits 808 nm beacon light, thus completing the coarse tracking process. When the beacon light enters the field of view of the fine tracking camera, the system automatically initiates the fine tracking process. After the completion of fine tracking, the drone transmits 1550 nm quantum light. 

As shown in Fig.1e, the 1550 nm optical beam is first chopped into optical pulses with a repetition rate of 10 MHz via the first Sagnac loop, followed by amplitude modulation using the second Sagnac interferometer. The amplitude-modulated signal is directed to the third Sagnac loop, subsequent to which it is split into signal light and LO light by a polarizing beam splitter (PBS). The signal light undergoes random phase modulation through the third phase modulator (PM) and is thereafter combined with the LO light using a second PBS. The combined optical signal is coupled into free space via a fiber collimator and transmitted after being expanded by a beam expander. Upon arriving at the UGV terminal, the combined light is split using a 90:10 beam splitter: the 10$\%$ optical branch is employed for the measurement of Stokes parameter $\hat{s}_{1}$, while the 90$\%$  branch is routed to a 50:50 beam splitter, with heterodyne detection subsequently performed to realize the measurements of Stokes parameters  $\hat{s}_{2}$  and  $\hat{s}_{3}$\cite{zheng2023experimental}.

\subsection{Experiments of drone hovering at different altitudes}

In establishing a hundred-meter-class QKD link between the drone hovering at different altitudes and fixed ground nodes, we set the drone's hovering heights to 25 m, 50 m, and 75 m, respectively, with 100 m distance from the stationary UGV. For more effective comparison, we set up an urban free-space quantum communication link between two fixed ground points 100 m apart. When the drone operates in hover mode, its sustained jitter induces phase fluctuations in the fast and slow axes of the encoding-dedicated fiber, thereby leading to continuous variations in the polarization state during encoded signal transmission\cite{novik2025time,braband2025fast}. To address this issue, this study adopts an encoder phase compensation method based on Stokes parameters, enabling real-time compensation for polarization state variations induced by phase perturbations\cite{2025Free}. During the preliminary system operation and performance verification phase prior to the experiment, we collected the encoded polarization state parameters corresponding to the $\hat{s}_{2}$  and  $\hat{s}_{3}$ components. Their distribution characteristics comply with the requirement of the Gaussian distribution, as illustrated in Fig.2a.

After the QKD transceiver equipment operates stably for half an hour, the experiment is initiated with a single test duration of approximately ten minutes. The system's channel loss under different operating conditions is as follows: for the free-space link between two fixed ground nodes separated by 100 m, the channel attenuation was 1.045dB; when the drone maintains a horizontal distance of 100 m from the UGV and performs a vertical takeoff, it hovers at heights of 25 m, 50 m, and 75 m respectively, with the corresponding channel losses being 0.955 dB, 1.369 dB, and 0.741 dB (Fig.2c,e,g,i). 
\begin{center}
	\includegraphics{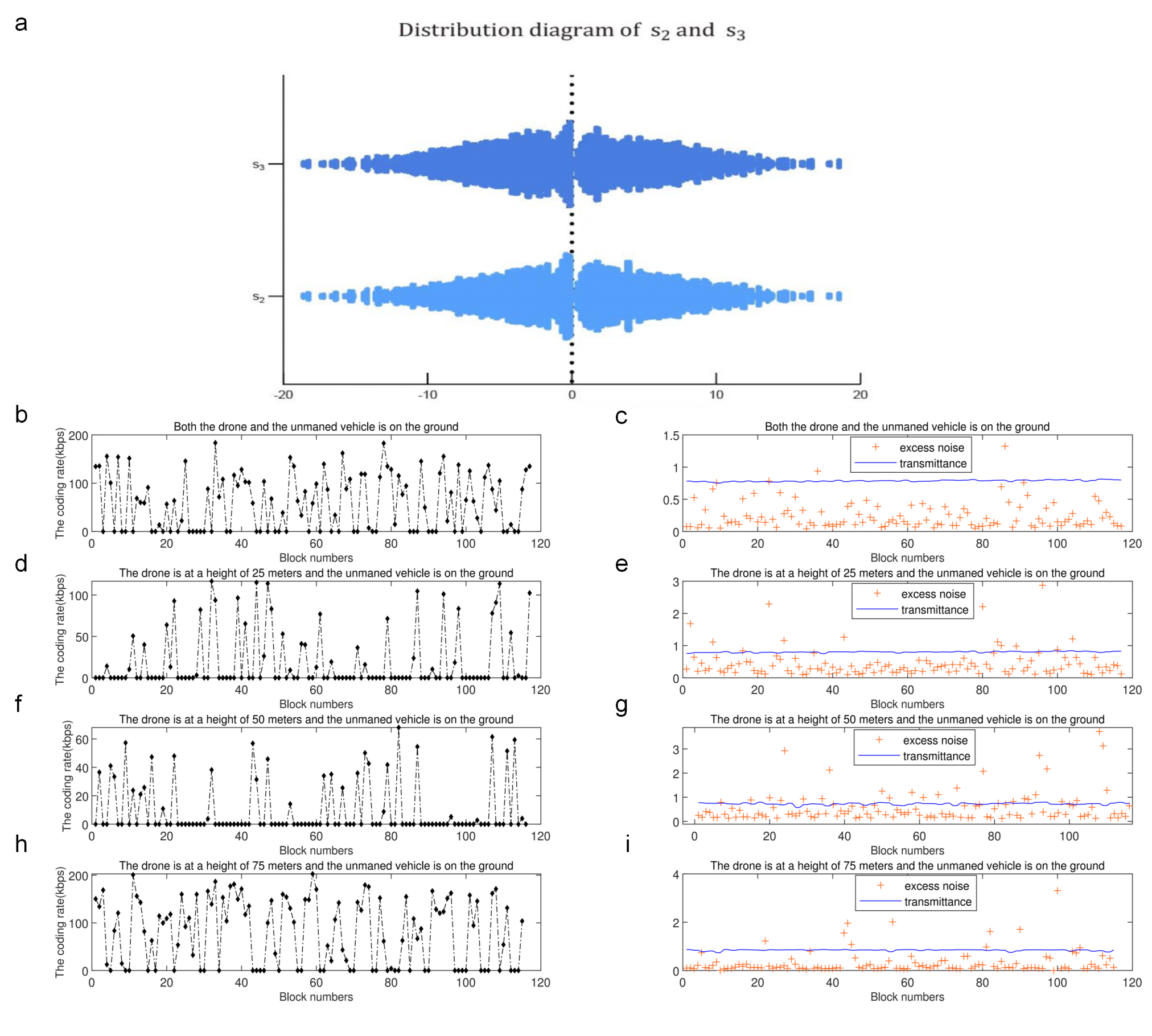}\\[8pt]
	\parbox[c]{15.0cm}{\footnotesize\textbf {Fig.~2.}   
		Experiments of drone hovering at different altitudes. (a)Distribution of components. (b)The key generation rate and transmittance between the drone hovering at different altitudes and the ground unmanned vehicle.\label{fig2}}
\end{center}
Processing the experimental data yields an average secure key rate of 58.69 kbps between two fixed ground points 100 m apart, with corresponding code generation rates of 19.14 kbps, 9.62 kbps, and 79.48 kbps for drone hovering altitudes of 25 m, 50 m, and 75 m relative to the stationary UGV, respectively(Fig.2b,d,f,h).
Experimental results demonstrate that the key rate between the drone and the UGV exhibits an inverse relationship with altitude within the 0-50 m range, reaching its peak at a hovering height of 75 m.
Analysis of the transmittance characteristics in Fig.2b-i yields a clear trend. For altitudes below 50 m, transmittance correlates inversely with height. In contrast, the transmittance at 75 m is significantly higher than the ground-level value. This discrepancy may be attributed to the dynamic changes in turbulence intensity at different altitudes and the influence of the zenith angle\cite{zou2018vertical,2017Satellite,2019Satellite,0Feasibility}.

\subsection{Experiments under relative motion}
In the experimental scenario where the drone is in flight, we observed that the system failed to achieve effective generation of quantum keys after the optical path was constructed. After analysis, this phenomenon is attributed to the polarization coordinate offset caused by the drone in the flying state. To address this technical bottleneck, we adopted an adaptive polarization calibration technology, which successfully resolved the problem of polarization coordinate deviation and thereby ensured the smooth progress of the experiment. More importantly, in the relative-motion experiments between drones and fixed ground nodes, Doppler shift and the pointing errors of PAT system constitute two critical technical challenges that must be thoroughly addressed\cite{li2025realizing}. For scenarios where the relative motion speed is relatively high and the Doppler shift effect is non-negligible, the countermeasures adopted by the experimental system are as follows: relying on a high-precision time synchronization system with a synchronization accuracy of 200 ps to ensure temporal synchronization, and applying a modulation voltage with a Gaussian distribution to the PM in the first Sagnac loop, so as to achieve effective compensation for the Doppler shift effect. These issues can ultimately be mitigated through the precise control of motion speed and rational trajectory planning. Based on preliminary experiments and in conjunction with the inherent performance constraints of the drone, we ultimately determined to regulate the drone flight speed to 1 m/s and adopt a linear path as its movement trajectory. This motion pattern could effectively simulate two practical scenarios: one with relative velocity between the drone and UGV, and another involving the slow variation of the angle between the drones velocity direction and the communication link (ranging from 87.7° to 92.3°).

To investigate QKD performance under relative motion between the drone and the UGV, we conducted experiments on a clear afternoon. In the first experiment, a static ground-based quantum communication link was established, with two fixed ground nodes deployed at its two ends and separated by a distance of 100 m. Experimental measurements demonstrated that the channel loss of this static link was 0.979 dB, and its average key generation rate reached 134.03 kbps(Fig.3a-b). In the second, a QKD link consisting of a hovering drone and a stationary UGV was established. The drone executed a vertical takeoff from a position with a horizontal distance of 100 m from the UGV, and then hovered stably at an altitude of 25 m to conduct QKD communication experiments with the stationary UGV. Experimental results demonstrated that the channel loss of this link was 1.093 dB, and its average key generation rate was 101.347 kbps(Fig.3c-d). In the third, a QKD link between a dynamically flying drone and a stationary UGV was established. In the experiment, the drone was maneuvered to perform a linear round-trip flight with a total travel distance of 10 m at a constant speed of 1 m/s, within a horizontal plane maintained at an altitude of 25 m. Meanwhile, with the stationary UGV located at a horizontal distance of 100 m serving as the fixed communication terminal, a quantum communication link was established between the drone and the UGV for the purpose of key negotiation and distribution. Experimental measurements indicated that the channel loss of this dynamic link was 0.959 dB, and its average key generation rate reached 70.94 kbps(Fig.3e-f).
\begin{center}
	\includegraphics{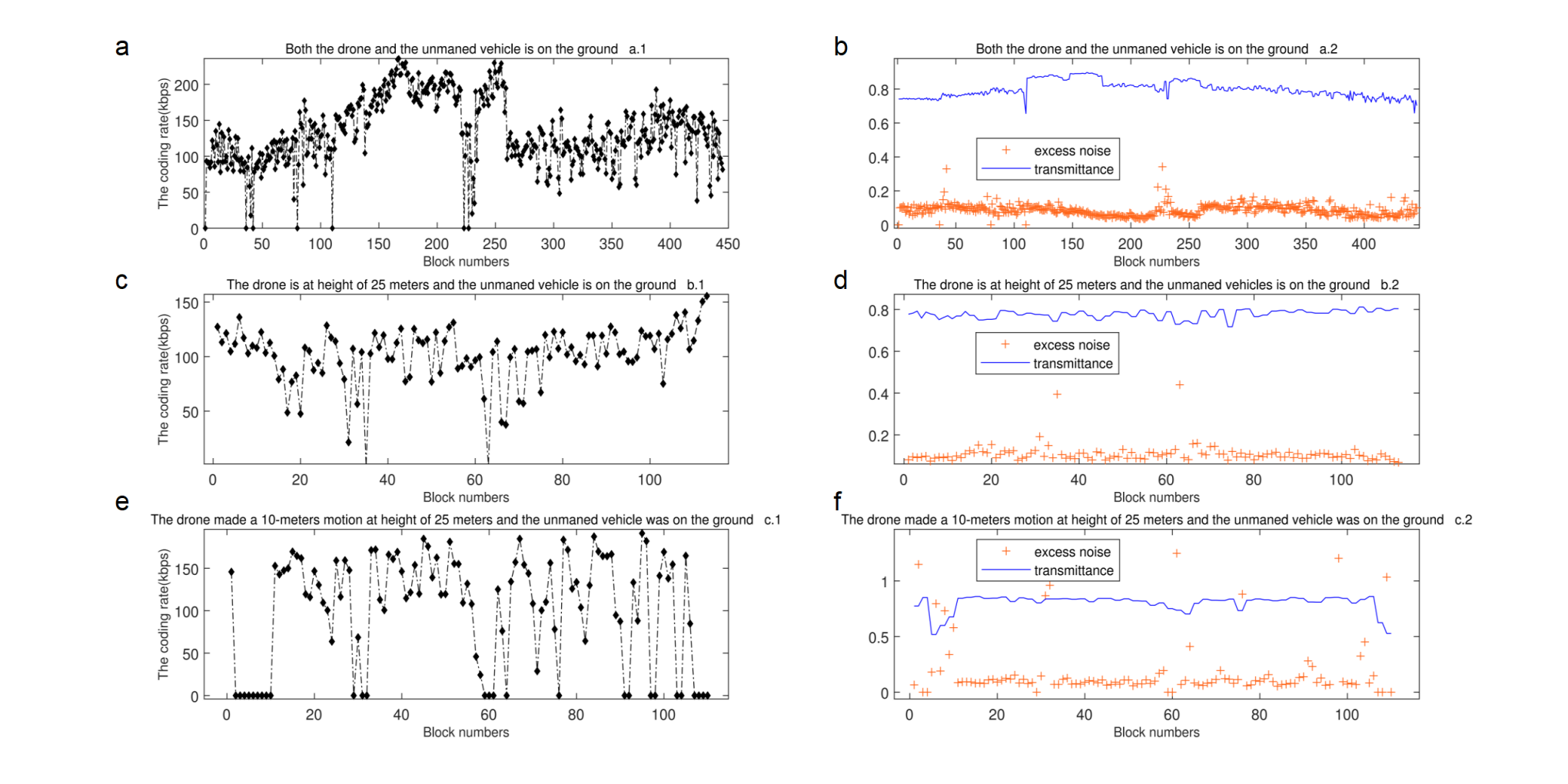}\\[8pt]
	\parbox[c]{15.0cm}{\footnotesize\textbf {Fig.~3.}  
		The key generation rate and transmittance between drone in different motion states and ground unmanned vehicles.\label{fig3}}
\end{center}

Our experiments reveal that a subset of single-round secure key rate between a drone in back-and-forth flight and a stationary UGV exceeds that of a hovering drone under identical conditions, yet remains lower than the average key rate of a quantum link between two fixed ground points at the same distance. Findings from the first experimental campaign elucidate the underlying mechanism governing the maximum key rate achieved between two fixed ground nodes. Evidenced by the data presented in Fig.3, the third-type link outperforms the second-type link in transmittance performance. Notably, the phase compensation values in the third scenario exhibit superior stability compared to those in the second scenario, which suggests, to a certain extent, that the ambient environmental conditions surrounding the drone in the third scenario are more stable than those in the second scenario. 
Nevertheless, the detrimental effects of relative motion compromise the PAT performance in this scenario, rendering its performance inferior to that of the second scenario, and thereby resulting in fewer key packets generated over the same time interval\cite{najafi2020statistical}. These findings provide crucial insights for constructing drone-based QKD links over kilometer-scale transmission distances.

\subsection{Kilometer-scale experiments}

In our exploration of kilometer-scale QKD in urban free space, we first established a fixed point-to-point QKD link spanning 1.2 km on a clear winter morning. The measured transmissivity of the communication link in this study was 0.483, and this metric demonstrated that the link had the technical feasibility to implement QKD. Subsequently, further experimental verification work pertaining to QKD would be carried out on the basis of this link.The experiment yielded an average key rate of 4.27 kbps(Fig.4d). Experimental results demonstrate that the PAT system employed can achieve real-time alignment of optical paths at kilometer-scale distances, and prove that QKD systems can still generate quantum keys despite the influence of kilometer-scale atmospheric turbulence. This provides technical support for exploring kilometer-scale QKD based on drone. During the experiment, we collected data on the PAT performance(Fig.4b). The achieved tracking accuracy at the UGV terminal is presented in Fig.4c. 

Secondly, owing to the constraints of the experimental space, we set the drone's hovering altitude to 25 m and established a quantum communication link with a fixed point located 1.2 km horizontally, as shown in Fig.4a. The experiment yielded an average key rate of 2.76 kbps, The relevant experimental details are further described, with results presented in Fig.4d.
\begin{center}
	\includegraphics{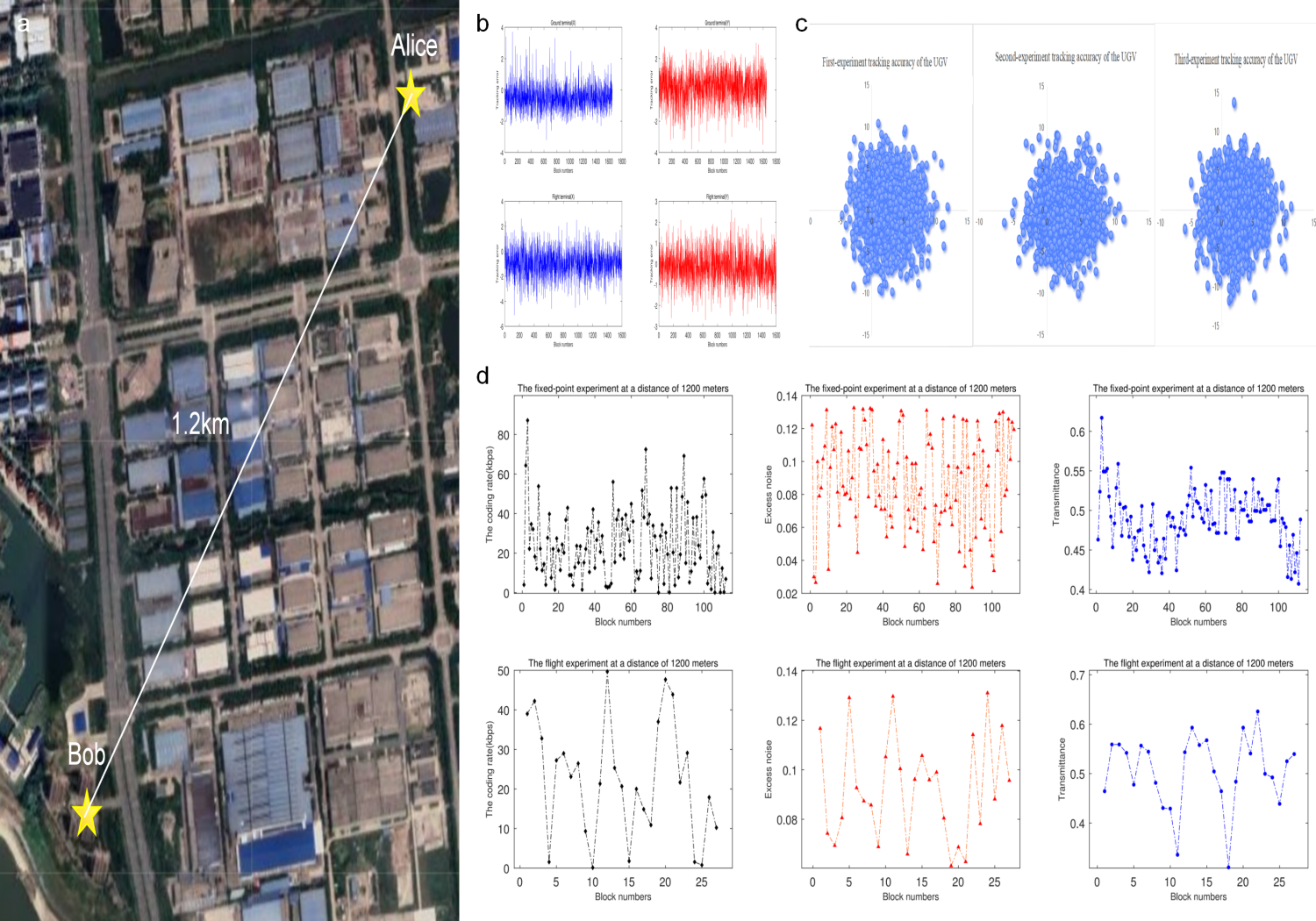}\\[8pt]
	\parbox[c]{15.0cm}{\footnotesize\textbf {Fig.~4.}  
		kilometer-level QKD experiments. (a)Top-down view of the experimental configuration in Wuhu, China, with the drone and UGV spaced 1.2 km apart. (b)The PAT errors in kilometer-level QKD links. (c)The PAT accuracy of the UGV. (d)The key generation rate and transmittance of kilometer-level QKD links. \label{fig4}}
\end{center}
The loss value of the kilometer-scale communication link is 3.468 dB, which is significantly higher than that of the other links. This result indicates that the kilometer-scale link is more significantly affected by atmospheric turbulence\cite{ghalaii2022quantum,sangeetha2024impact}.

\section{ DISCUSSION}\label{sec3}

We successfully demonstrated daytime free-space QKD from a drone to a fixed point over a distance of 1.2 km, with the communication range extendable to 2.4 km through trusted relays. Experimental findings reveal that, under identical horizontal distances, the key rate does not necessarily increase with drone altitude - specifically, when the altitude is below 50 m, the average secure key rate exhibits an inverse relationship with height. Furthermore, compared to flight operations, a hovering drone establishes more stable links with UGV, whereas a drone flying at low speed along a straight path demonstrates higher key generation rates when communicating with fixed points.

Compared to previous experiments, our experiment achieves longer distances and higher secure key rates. Our contribution lies in the realization of QKD under conditions including the drone hovering at different heights, different motion states of the two communication parties, and communication distances exceeding the kilometer level. This experiment holds significant practical value, marking a major step forward in achieving dynamic key distribution, securing mobile communications, and supporting rapid network deployment. It successfully fills the technical gap in establishing highly secure communications on mobile drone platforms using CV-QKD technology. Future research should be further improved: optimizing ATP system performance to shorten tracking and aiming time, enhancing link stability and extending transmission distance, introducing adaptive optics to mitigate adverse effects of atmospheric turbulence, thereby improving the secure key rate.

\section{ METHODS}\label{sec4}

\subsection{Description of the GMCS CV-QKD protocol}

In the  QKD system of drone, we selected the GMCS CV-QKD protocol based on Stokes parameter encoding. This protocol is suitable for compact devices \cite{zheng2023experimental} and exhibits robustness against external environmental noise\cite{2025Free}. The continuous polarization state modulation mechanism of the protocol is as follows: The drone transmitter randomly generates two independent parameters ${U}_{1}$ and ${U}_{2}$ (both following a uniform distribution over the interval $\left[ 0,1\right]$), which respectively regulate the phase modulators embedded in the second and third-stage Sagnac Interferometers  to achieve programmable control of the polarization state(Fig.1c).
The optical signal is modulated by the first Sagnac Interferometer and converted into a pulsed optical signal. It is then adjusted to a 45° linear polarization state via a polarization-maintaining faraday Rotator(PMFR), denoted as $\rvert\psi_{in} \rangle$. 
\begin{align}
	\rvert\psi_{in} \rangle=\frac{a_{LO}}{\sqrt{2}}\left( \rvert H\rangle +\rvert V\rangle\right) \label{eq4} 
\end{align}
Upon entering the second Sagnac Loop, $\rvert\psi_{in} \rangle $ is split into two orthogonal polarization components (horizontal $\rvert\ H \rangle $ and vertical $\rvert\ V \rangle $) by a PBS, PM1 applies a relative phase shift controlled by parameter ${U}_{1}$ to the horizontal component, and the resulting polarization state is denoted as $\rvert\psi_{1} \rangle $.
\begin{align}
	\rvert\psi_{1} \rangle=\frac{a_{LO}}{\sqrt{2}}\left( e^{i\phi_{1}}\rvert H\rangle +\rvert V\rangle\right)=\frac{a_{LO}}{\sqrt{2}}\begin{bmatrix} e^{i\phi_{1}} \\ 1 \end{bmatrix} \label{eq5} 
\end{align}
$\rvert\psi_{1} \rangle$ undergoes a linear transformation of the polarization state through the optical path system composed of the second Sagnac Loop and PBS, yielding the polarization state $\rvert\psi_{2} \rangle$ after emission.
\begin{align}
	\rvert\psi_{2} \rangle=\frac{a_{LO}}{\sqrt{2}}\begin{bmatrix} 1 \\ -e^{i\phi_{1}} \end{bmatrix} \label{eq6} 
\end{align}
Subsequently, $\rvert\psi_{2} \rangle$ undergoes polarization direction calibration via the PMFR again to obtain the pre-modulated state $\rvert\psi_{3} \rangle$.
\begin{align}
	\rvert\psi_{3} \rangle=\frac{a_{LO}}{2}\begin{bmatrix} 1-e^{i\phi_{1}} \\ -\left( 1+e^{i\phi_{1}}\right)  \end{bmatrix} \label{eq7} 
\end{align}
$\rvert\psi_{3} \rangle$then enters the third-stage SL and is subjected to a secondary phase modulation by the embedded PM (controlled by parameter ${U}_{2}$), resulting in the intermediate polarization state $\rvert\psi_{4} \rangle$.
\begin{align}
	\rvert\psi_{4} \rangle=\frac{a_{LO}}{2}\begin{bmatrix} \left( 1-e^{i\phi_{1}} \right)e^{i\phi_{2}} \\ -\left( 1+e^{i\phi_{1}}\right)  \end{bmatrix} \label{eq8} 
\end{align}
Finally, $\rvert\psi_{4} \rangle$ is combined by the third-stage PBS and outputs the target modulated polarization state $\rvert\psi_{5} \rangle$.
\begin{align}
	\rvert\psi_{5} \rangle=\frac{a_{LO}}{2}\begin{bmatrix} -\left( 1+e^{i\phi_{1}}\right) \\ -\left( 1-e^{i\phi_{1}} \right)e^{i\phi_{2}}   \end{bmatrix} \label{eq9} 
\end{align}

The parameters $\hat{s}_{2}$ and $\hat{s}_{3}$ of this polarization state satisfy the conditions :
\begin{align}
	\hat{s}_{2} &= x_{1} = \sqrt{-2V_{1}\ln\left( U_{2}\right) } \cdot \cos\left( 2\pi U_{1}\right) \nonumber \\
	\hat{s}_{3} &= p_{1} = \sqrt{-2V_{1}\ln\left( U_{2}\right) } \cdot \sin\left( 2\pi U_{1}\right) \label{eq1}
\end{align}

After reception at the UGV receiver, the signal is detected using two homodyne detectors via heterodyne detection methodology. Subsequently, parameter estimation, error correction, and privacy amplification are performed to generate the final key. The secure key rate formula is given by\cite{leverrier2010finite}:
\begin{equation}
	K = f\dfrac{n}{N}\left[ \beta I \left(  A:B\right)-S_{\epsilon_{PE}}\left( B:E\right) -\Delta n\right] \label{eq2}
\end{equation}
Here, $f$ denotes the repetition rate, $n$represents the number of key-contributing bits, $N$ indicates the total number of packets, $ \beta $corresponds to the reverse negotiation efficiency, $I \left(A:B\right)$signifies the mutual information between the communicating parties, $S_{\epsilon_{PE}}\left( B:E\right)$ represents the maximum value compatible with the statistics provided by parameter estimation except for probability $\epsilon_{PE}$, and $\Delta n$ is associated with privacy amplification.

\subsection{Pointing, acquisition, and tracking system}

The system employs a vibration-isolation platform to suppress high-frequency vibrations and implements both coarse and fine tracking using two closed-loop control systems to enable scanning, acquisition, and tracking functions. To achieve highly stable reception of 1550 nm quantum signal light, the optical system employs a high-precision common-path design that integrates the quantum optical path, the classical communication optical path, and the fine-tracking detection optical path. Moreover, both coarse and fine tracking use the same camera model with a pixel array of $2048\times 2048$. The coarse-tracking camera initially uses a $1024\times1024$ pixel region, while the fine-tracking camera is initially set to $512\times512$ pixels, with each pixel measuring 5.5 $\mu$m. During fine tracking, the camera performs a windowed readout based on the position of the fine-tracking spot, dynamically adjusting the readout area to $512\times512$, $256\times256$, $128\times128$, or $64\times64$ pixels to improve tracking performance.  Meanwhile, the exposure time is controlled on a millisecond-to-second scale according to the spot image information, enabling more accurate calculation of the spot centroid. The dual-beacon design of the PAT system reduces link establishment latency and enables rapid establishment of the communication link\cite{chun2017handheld}.

As shown in Fig.5a, in the drone-mounted setup, the coarse-tracking camera detects the optical spot in the 808 nm band. The data processing module then calculates the centroid position of the spot, and the control circuit drives the servo gimbal to achieve coarse tracking, with a pointing accuracy of 323 $\mu$rad. Subsequently, the fine-tracking camera captures the spot in the 808 nm band. Based on its centroid position, the control circuit drives the fast-steering mirror to perform fine tracking.The dynamic tracking accuracy measured at 100 m reaches 38 $\mu$rad.  
\begin{center}
	\includegraphics{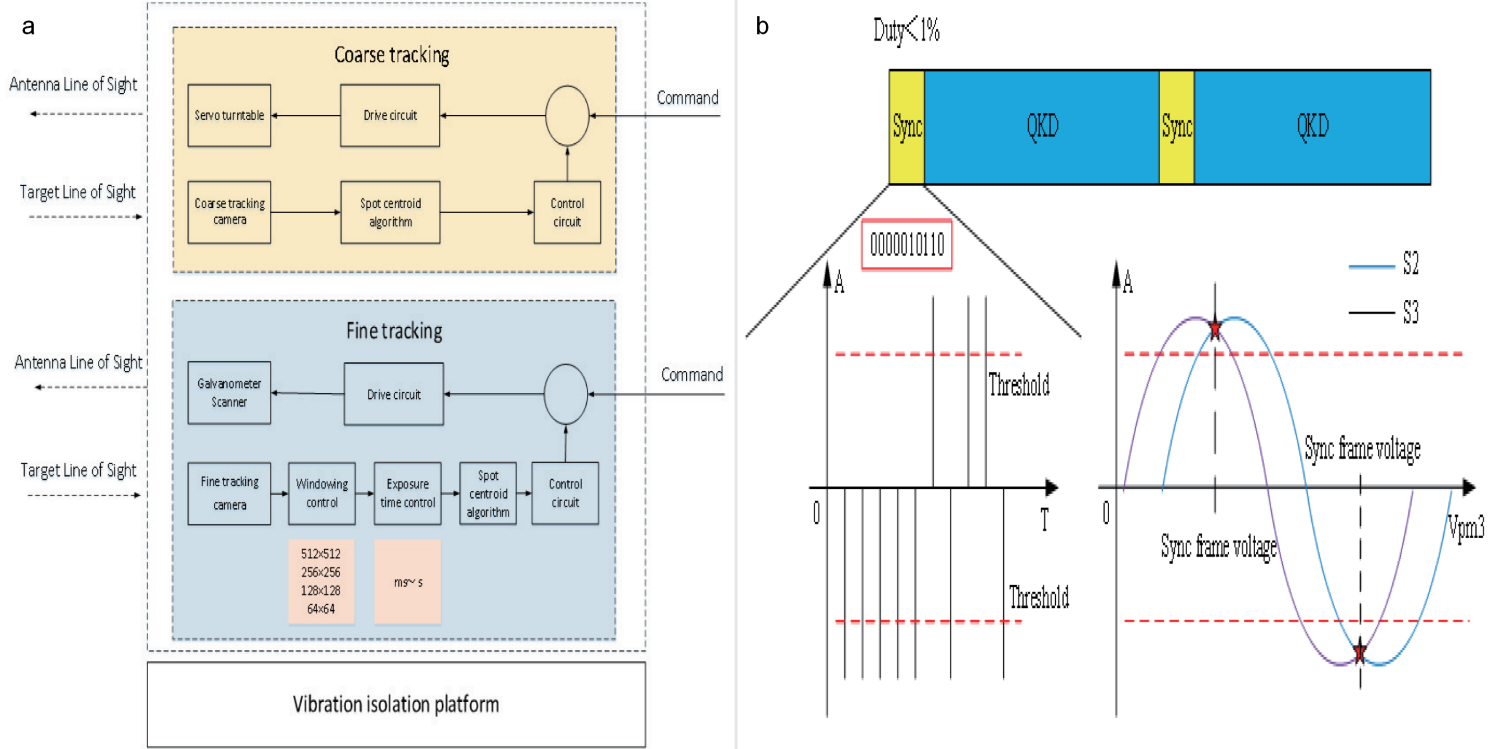}\\[8pt]
	\parbox[c]{15.0cm}{\footnotesize\textbf {Fig.~5.}  
		Schematic Diagram of the Methodology. (a)Schematic diagram of the scanning, acquisition and tracking implementation. (b)PM synchronization frame voltage initialization scanning process.\label{fig5}}
\end{center}

\subsubsection{Phase modulator-based synchronization method}

Using a quantum-light laser and homodyne detectors, the system achieves high-efficiency synchronization between Alice and Bob without any extra hardware. Implementation details: the entire transmission is divided into a synchronization phase followed by the QKD phase. The synchronization phase consists of a special 10-bit pattern (0000010110), where 0 denotes a negative electrical level and 1 a positive level; the duty cycle is $<$1$\%$ . The timing diagram is shown in Fig.5b.

At the receiver, the FPGA opens a synchronization window. If the detection results inside this window match the transmitted sync pattern, Alice and Bob are declared synchronized and proceed to key distribution. To prevent this sync pattern from being mistaken for data during the subsequent QKD stage and thus avoiding false re-synchronization the amplitude threshold for the fixed pattern is deliberately set higher than the signal amplitude used during the actual key-distribution phase.
\begin{align}
	\hat{s}_{2} &= a_{LO}\sin\left(\phi_{1} \right) \sin\left( \phi_{2}\right)   
	\nonumber \\
	\hat{s}_{3} &= a_{LO}\sin\left(\phi_{1} \right) \cos\left( \phi_{2}\right) \label{eq3}
\end{align}

During the synchronization-frame voltage determination, the drive voltages of phase modulators $ {PM}_{2}$ and ${PM}_{3}$ are adjusted as follows: ${PM}_{2}$  is fixed so that $\sin\left(\phi_{1} \right) $ is maximized. Owing to slow phase drifts caused by the environment, ${PM}_{3}$ is swept over $0-2\pi$ to locate the optimum sync-frame voltage at which both ${s}_{2}$ and ${s}_{3}$ reach their preset thresholds. During transmission, the sync-frame pulses are sent and received using the sync-frame voltage obtained from the preceding scan; an updated PM sync-frame scan is executed immediately before each transmission. This approach ensures reliable detection of the synchronization signal, clearly separates the synchronization and data-communication phases, and enhances the overall stability and accuracy of the system. Throughout the entire drone flight experiment, the sync light caused virtually no anomalies that would disturb normal system operation.

\addcontentsline{toc}{chapter}{Appendix A: Appendix section heading}
\section*{FUNDING \label{app1}}
This work was supported by the Henan Science and Technology Major Project of the Department of Science(241100210400).

\section*{AUTHOR CONTRIBUTIONS \label{app2}}

 C.Z. and Y.Y.Z. supervised and conceived this work. 
 C.Z., H.L., D.W.L., F.X.X., Q.F.Z., H.S., and S.B.T. designed the research framework and carried out the numerical simulations.
 Y.Y.Z., Q.F.Z., Y.Z., Y.M.Z., D.Y.Z., P.W., and H.W. performed the experiments and analyzed the resulting data.
 C.Z., Y.Y.Z., Y.Z., T.L., and P.W. processed the experimental data and contributed to the preparation of the original manuscript.
\bibliographystyle{unsrt}
\bibliography{nsr_sample}
\end{document}